# Quantum light generation with ultra-high spatial resolution in 2D semiconductors via ultra-low energy electron irradiation

Ajit Kumar Dash[1], Sharad Kumar Yadav[1], Sebastien Roux[2], Manavendra Pratap Singh[3], Kenji Watanabe[4], Takashi Taniguchi[5], Akshay Naik[3], Cedric Robert[2], Xavier Marie[2], Akshay Singh[1*]

[1]*Department of Physics, Indian Institute of Science, Bengaluru, Karnataka 560012, India*

[2]*Université de Toulouse, INSA-CNRS-UPS, LPCNO, 135 Avenue de Rangueil, 31077 Toulouse, France*

[3]*Centre for Nanoscience and Engineering, Indian Institute of Science, Bengaluru, Karnataka 560012, India*

[4]*Research Center for Functional Materials, National Institute for Materials Science, Japan*

[5]*International Center for Materials Nanoarchitectonics, National Institute for Materials Science, Japan*

*Corresponding author: aksy@iisc.ac.in

**Abstract:**

Single photon emitters (SPEs) are building blocks of quantum technologies. Defect engineering of 2D materials is ideal to fabricate SPEs, wherein spatially deterministic and quality-preserving fabrication methods are critical for integration into quantum devices and cavities. Existing methods use combination of strain and electron irradiation, or ion irradiation, which make fabrication complex, and limited by surrounding lattice damage. Here, we utilise only ultra-low energy electron beam irradiation (5 keV) to create dilute defect density in hBN-encapsulated monolayer $MoS_2$, with ultra-high spatial resolution (< 50 nm, extendable to 10 nm). Cryogenic photoluminescence spectra exhibit sharp defect peaks, following power-law for finite density of single defects, and characteristic Zeeman splitting for $MoS_2$ defect complexes. The sharp peaks have low spectral jitter (< 200 $\mu$eV), and are tuneable with gate-voltage and electron beam energy. Use of low-momentum electron irradiation, ease of processing, and high spatial resolution, will disrupt deterministic creation of high-quality SPEs.

**Introduction:**

Quantum technologies (quantum computing, sensors and communication) are emerging as a transformative force in this information age, and at the core, lies the critical need for reliable sources of single photons, known as single photon emitters (SPEs)[1–5]. Two-dimensional (2D) materials have emerged as compelling candidates for SPEs due to their distinctive electronic and optical characteristics[6–9]. Notably, 2D transition metal dichalcogenides (TMDs) offer versatility in material processing and exhibit a wide array of emission wavelengths, operation temperatures, charge tunability, and polarisation properties[6,7,10–13]. SPEs in 2D materials represent a significant leap over their 3D counterparts (nitrogen vacancy centres, epitaxial quantum dots) regarding integration and compatibility with photonic structures, positioning defects near the surface for more efficient interaction with fields in sensing applications, and low-complexity synthesis enabling scalable and cost-effective fabrication.

The deterministic creation of SPEs in 2D materials is necessary for integration into practical devices due to constraints on cavity coupling and electrical device integration. Highly spatially deterministic SPE creation (< 50 nm) will enable SPE coupling with dielectric and plasmonic cavities having very small mode volumes for high Purcell enhancement, as well as specially designed cavities for enhancing out-coupling in the direction of collection optics[14–16]. Studies of dipole-dipole interactions can also be enabled in solid-state systems by creating SPEs in close proximity (5-50 nm). Various techniques have been employed for deterministic creation of SPEs, each with advantages and limitations. High energy $He^+$ ion irradiation[17], UV light irradiation[18], strain engineering using nanopillars[19,20] and atomic force microscopy (AFM) nanoindentation[21], combined strain and defect engineering[22,23], and combined strain and chemical treatment[24] have been employed successfully for SPE fabrication in atomically thin TMDs. Ion irradiation can deterministically create SPEs in 2DMs, however, surrounding lattice damage can happen due to the high energy of ion beams[25–27]. UV light irradiation is relatively simple, but lacks the precision (spatial resolution) offered by electron or ion beam irradiation. Strain engineering techniques and chemical approaches face severe challenges on multiple defect creation at same site, reproducibility, defect passivation and environmental sensitivity. Alternatively, electron beam (e-beam) irradiation can be utilised to deterministically create defects.

The momentum of electrons is very low compared to ions (nearly 7,400 times for the same accelerating voltage). Hence, using low-energy e-beam to create defect emission centres is expected to cause less damage to the surrounding lattice than ion-beam irradiation, which may result in superior SPE performance. In our earlier work[28], we utilized ultra-low energy e-beam irradiation (≤5 kV) to achieve highly controlled defect formation. Thus, the utilisation of ultra-low energy e-beam irradiation for SPE generation in TMDs offers unparalleled control and precision, however with no experimental demonstrations so far.

In this work, we utilise ultra-low energy (5 kV) e-beam for creating dilute density of defects in monolayer (ML) $MoS_2$, followed by quick encapsulation between hBN layers. The e-beam irradiation results in creation of ultra-sharp (linewidths < 1 meV), bound exciton peaks 100 – 300 meV below the neutral exciton peak (in photoluminescence (PL) spectra). Further, we confirmed that the individual sharp peaks follow power-law saturation for two-level system of single defects, and have high spectral stability over time (spectral jitter < 200 $\mu$eV). We found that these sharp peaks have majority of PL in the zero-phonon line (ZPL), with a minor contribution from phonon side band (PSB) at the lower energy side. We also discuss the defect origin of the sharp peaks, supported by magneto-optics measurements, and attribute them to be SPEs. Further, we found that the SPEs are tuneable with electron beam energy and gate voltage. At last, we showed spatially deterministic placement of SPEs with resolution < 50 nm by e-beam lithography-based irradiation. Our research could disrupt the fabrication methodology for SPEs, laying the foundation for the next generation of quantum technologies.

## Results:

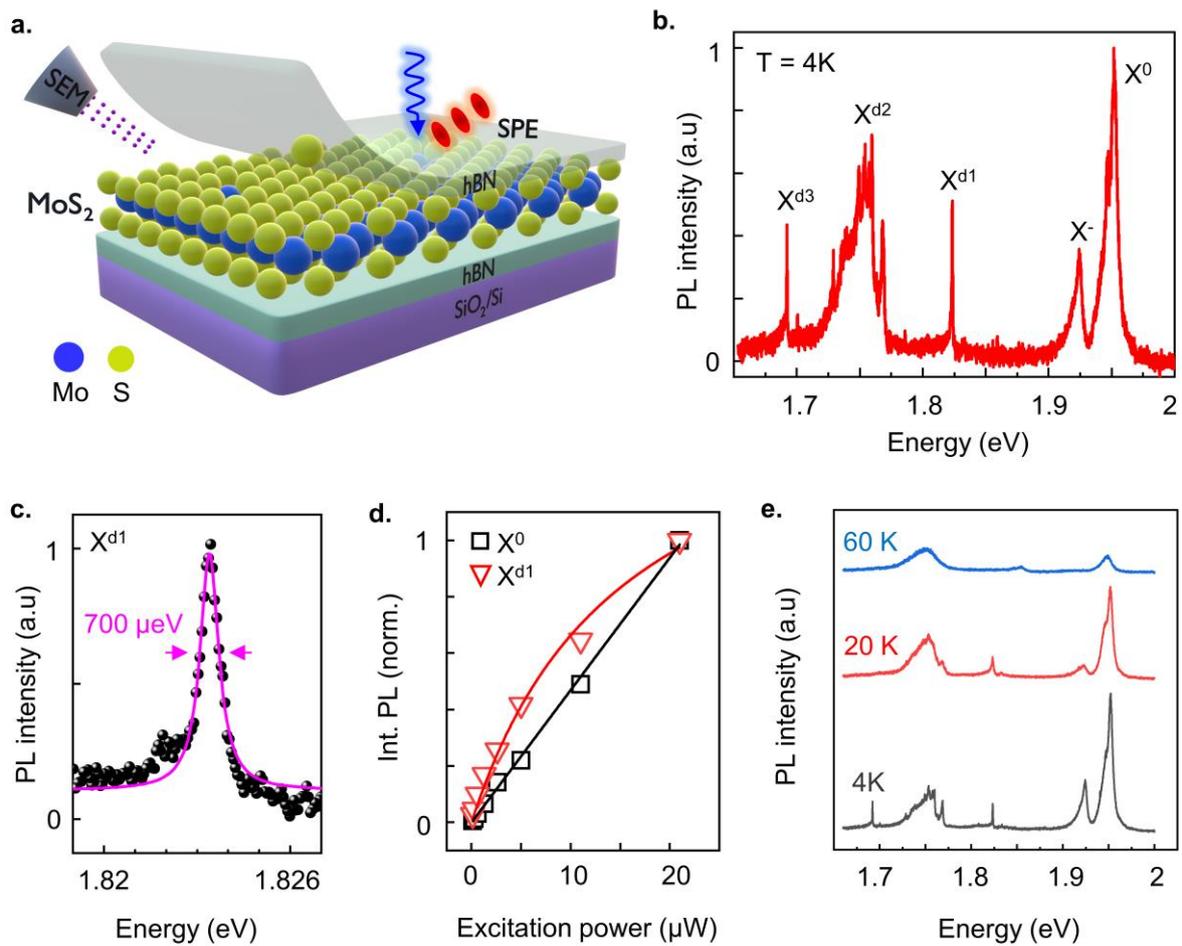

**Figure 1: Ultra-narrow linewidth, and bound photoluminescence (PL) peaks induced by electron beam (e-beam) irradiation. (a)** Schematics showing defect formation in monolayer (ML) TMDs by e-beam irradiation, followed by hBN encapsulation, and single photon generation from defect sites. **(b)** Representative cryogenic PL spectra of irradiated sample. $X^0$, $X^-$, and $X^d$ denote neutral, charged, and defect bound excitons, respectively. **(c)** Lorentzian fitting of $X^{d1}$ peak. **(d)** The power law fitting of $X^0$, $X^{d1}$ peaks. **(e)** Evolution of cryogenic PL with temperature.

**E-beam irradiation of ML MoS$_2$:**

MLs of MoS$_2$ were exfoliated onto 285 nm SiO$_2$-coated silicon substrate by PDMS-assisted blue tape technique (see Methods). Freshly exfoliated ML flake was raster scanned by e-beam of energy 5 keV, at a constant dose of $1.57 \times 10^{13}$ electrons per cm$^2$ (e$^-$/cm$^2$) in a scanning electron microscope (SEM) chamber (maintained at $3 \times 10^{-5}$ mbar vacuum). See Methods and Supplementary Information SI-I for details. To avoid environmental passivation of e-beam induced defects, the irradiated ML was encapsulated between hBN layers within an hour of irradiation by viscoelastic transfer (see Methods). The electron irradiation in SEM, and post-irradiation hBN encapsulation process is schematically presented in Figure 1(a). The typical thickness of hBN flakes used in this study ranges from 10 – 25 nm, as confirmed by AFM.

**Cryogenic PL characterization:**

To confirm the creation of optically active defects in the irradiated sample, cryogenic PL (T = 4 K) spectra were acquired using a 478 nm continuous laser at low excitation power (~ 0.6 μW). Representative PL spectrum of hBN encapsulated irradiated-ML $MoS_2$ depicted in Figure 1(b) corroborates the presence of spectrally sharp (FWHM ~ 0.4 – 2 meV) excitonic peaks ($X^d$), at energies 100 – 300 meV below neutral exciton ($X^0$) peak. The sharp peaks appear in three spectral regions, labelled as $X^{d3}$ (1.68 -1.71 eV), $X^{d2}$ (1.72 – 1.77 eV), and $X^{d1}$ (1.81 – 1.85 eV). Physical origin of these different spectral regimes is discussed later. The observed sharp excitonic peaks ($X^d$) are asymmetric, with a minor contribution from the phonon sideband (PSB) at the lower energy side (SI-IX). Lorentzian fitting of the $X^{d1}$ peaks is illustrated in Figure 1(c). Interestingly, the sharp peaks are only observed if the hBN encapsulation is performed within a few hours of electron irradiation (SI-III). Hence, the e-beam induced defects get quickly passivated by environmental gases and hydrocarbons, which is also evidenced by previous work on $MoS_2$ defects created by proton beam irradiation[29]. Further, we measured PL of an encapsulated electron irradiated sample after one year (SI-IV), and still observed sharp peaks, indicating the long-term environmental stability of defects after hBN encapsulation.

Apart from $X^0$ and $X^d$, a less prominent trion peak ($X^-$) appears alongside the neutral exciton peak, indicating slight doping of ML $MoS_2$. Further, the extracted homogenous and inhomogeneous linewidths of $X^0$ peak are nearly 2.1 and 2.2 meV, respectively (SI-II). The reduced inhomogeneous linewidth of $X^0$ peak after hBN encapsulation indicates high interface quality of the prepared samples[30]. Also, we note that around a few spots in the irradiated samples, both $X^0$ and $X^-$ peaks show fine splitting, varying from 5 meV to 10 meV (SI-II). This exciton splitting is not the focus of this work, but it may be attributed to inhomogeneous or anisotropic strains developed by defects. Further, annealing of hBN encapsulated irradiated-ML MoS$_2$ in inert atmosphere causes broadening of the sharp defect peaks[31] (SI-V). However, optimisation of annealing conditions may improve[32] defect emission properties.

**Bound nature of e-beam induced sharp peaks:**

Excitation laser power-dependent PL was performed to validate the bound nature of sharp $X^d$ peaks. The integrated intensity of $X^0$ and $X^{d1}$ peaks are plotted against laser power in Figure 1(d). Raw PL data for all powers is presented in SI-VI. The $X^d$ peaks emission shows saturation behaviour with laser power, a signature of defect-bound excitons with a low density of states. For a two-level system, the saturation powers ($P_{sat}$) for finite density of single defects can be obtained by using the equation,

$$I = k \frac{P}{P+P_{sat}} \quad (1)$$

Where $k$ is a fitting constant, and $P$ and $I$ are laser power and PL intensity, respectively. The saturation power for $X^d$ peaks varies between 5 μW and 15 μW (laser spot size ~ 1.3 μm), similar to quantum emitters found in MoS$_2$[33]. The sharp $X^d$ peaks show sublinear (power coefficient ~ 0.7) dependence with laser power, unlike the usual linear laser power dependence (power coefficient ~ 1) of delocalized neutral exciton ($X^0$) and trion ($X^-$) peaks, which confirms the defect-bound nature of $X^d$ peaks (power coefficients and analysis of all exciton peaks is presented in SI-VI). To further validate the defect nature of $X^d$ peaks and investigate their temperature stability, we performed temperature-dependent PL of irradiated samples (Figure 1(e)). With increasing temperatures, the sharp peaks broaden and reduce in intensity. The $X^d$ peaks exist up to 60 K temperature, consistent with the bound exciton model description[34]. At 4 K temperature, line shape of the sharp defect peaks are nearly Lorentzian, and becomes quasi-Lorentzian at higher temperatures (SI-IX). With increase in temperature, the fraction of ZPL decreases[33,35] and PSB contribution increases, causing the broadening of the sharp defect peaks (SI-IX).

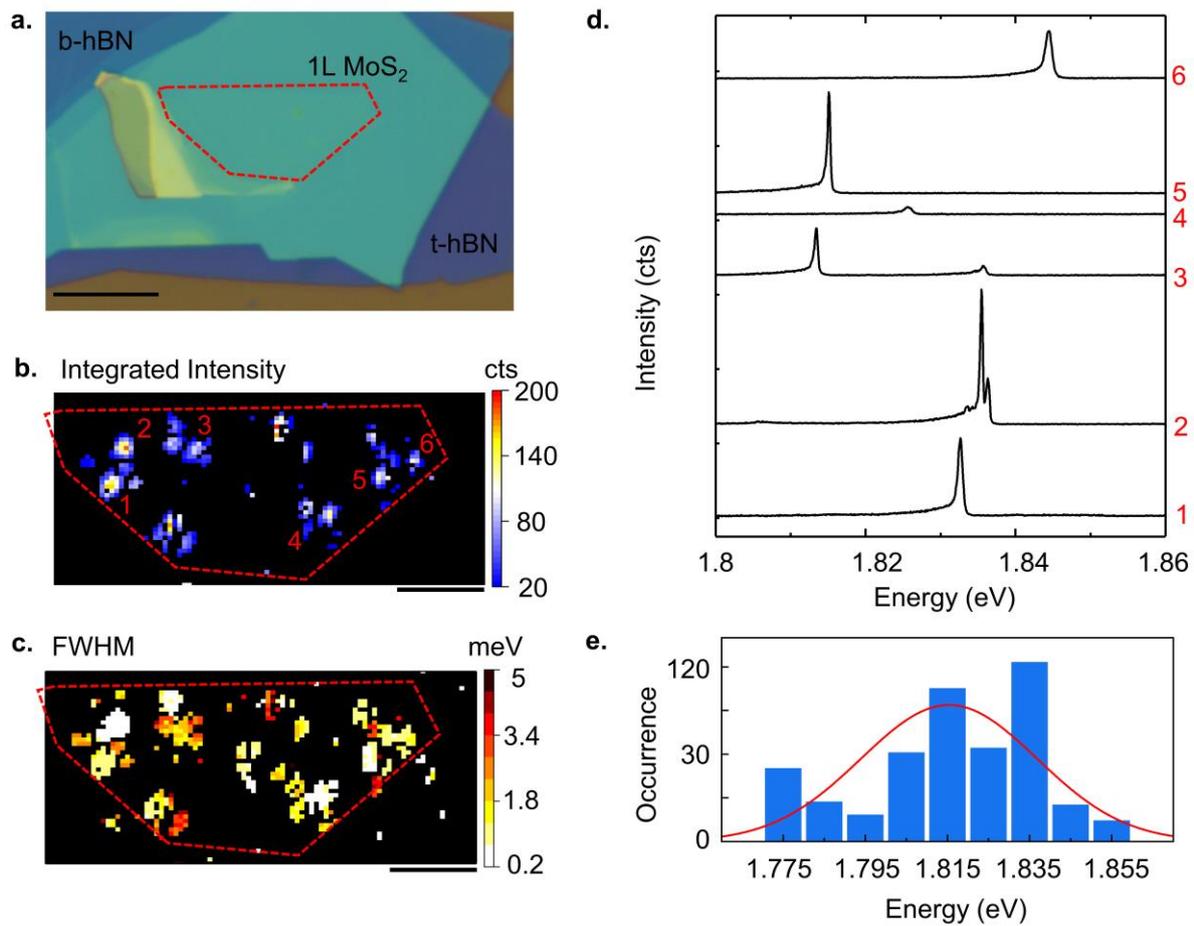

**Figure 2. Spatial and spectral distribution of sharp defect peaks prepared via large-area irradiation. (a)** Optical image of hBN encapsulated irradiated-ML MoS$_2$. The scale bar is 10 μm. Corresponding cryogenic spatial mapping of integrated PL intensity and linewidth of X$^{d1}$ peak are shown in **(b)** and **(c)**, respectively. The ML region is outlined using red-dashed lines. Scale bars correspond to 5 μm. **(d)** Cryogenic PL spectra of locations 1 – 6 marked in (b). **(e)** Histogram showing variation in the X$^{d1}$ peak position across the ML.

**Spatial and energy distribution of sharp defect peaks:**

Cryogenic PL mapping was carried out in a custom-built optical setup to study the spatial distribution of defect emission centres (see Methods). The optical image of hBN encapsulated irradiated-ML MoS$_2$, and corresponding spatial map of X$^{d1}$ peak integrated intensity is shown in Figure 2(a) and 2(b), respectively. Here, X$^{d1}$ peak is chosen for analysis because of its isolated nature from other sharp peaks. The X$^{d1}$ emission centres are localized around bright spots of the size ~ 500 nm (i. e., lateral resolution of the mapping setup), consistent with emissions from single defect sites. The map of linewidth of X$^{d}$ peaks across the sample is depicted in Figure 2(c). The majority of sharp peaks have line widths of less than 1 meV, suggesting the formation of highly localized and dilute density of defects. The raw PL maps and protocol for obtaining a filtered PL map are presented in SI-VII. Figure 2(d) presents raw PL spectra from locations (1 - 6) marked in Figure 2(b). The statistics of the spectral position of the $X^{d1}$ peak are plotted as a histogram in Figure 2(e). The sharp $X^{d1}$ peaks appear in a band of nearly 80 meV with the centre around 1.815 eV. The spread in spectral energy and linewidths of $X^{d1}$ peaks can be attributed to spatial variation in defect environment and/or inhomogeneous strain across the ML $MoS_2$.

**Time stability, gate-voltage dependent PL of sharp defect peaks:**

Energetically stable SPEs are ideal for quantum applications. The spectral jittering (average spectral shift of SPE peaks over time) measurement gives insights into the stability of SPEs, usually limited by charge fluctuations in the environment. Figure 3(a) shows the kinetic series of PL for over 2500 sec, with each PL spectra acquired for 5 sec. The SPE peak moves less than ± 200 μeV over time, indicating very high spectral stability of SPEs. We noticed that at high excitation power, the peak can jump by several meV in extended time-scale experiments, indicating modification of charge state of defects.

To further understand the charged nature of sharp defect peaks and effect of charge carrier doping, gate voltage dependent PL of hBN encapsulated irradiated-ML $MoS_2$ was performed. Schematics and optical image of the device is presented in SI-X. The evolution of cryogenic PL with gate voltage is shown as a surface plot in Figure 3(b). The raw PL spectra, plot of integrated intensity of PL peaks with applied gate voltage, and calculation of carrier density using capacitor model are presented in SI-X. By applying positive gate voltage (electron doping), we observed quenching of $X^0$ and increase of $X^-$ peak intensity. Also, $X^d$ and L peaks were quenched with electron doping. Interestingly, PL intensity change was minimal on hole doping side, consistent with earlier reports[36]. Hence, the intensity of sharp defect peaks are tuneable with the gate voltage.

**Characteristic valley Zeeman splitting of sharp defect peaks:**

In order to understand the origin and the spin properties of the sharp defect peaks, we performed magneto-PL with magnetic field oriented perpendicular to atomic layers (Faraday geometry). Circularly polarized PL ($\sigma^+$ and $\sigma^-$) was detected after excitation with linearly polarized laser. Raw spectra of $\sigma^+$ and σ⁻ emitted intensity of $X^{d1}$ peak at 0 T, 6 T, and -6 T magnetic field are shown in SI-XI. Data are fitted with a Lorentzian curve, from which the intensity ($I^\sigma$) and position ($E^\sigma$) are extracted. The degree of circular polarization (DCP = $(I^{\sigma^+} - I^{\sigma^-})/(I^{\sigma^+} + I^{\sigma^-})$) and energy splitting ($\Delta = E^{\sigma^+} - E^{\sigma^-}$) are deduced as a function of magnetic field in Figure 3(c) and 3(d), respectively. The experimental g-factor can be calculated from $\Delta = g\mu_B B$, where $\mu_B$ is the Bohr magneton. The corresponding g-factor of $X^{d1}$ peak obtained from slope ($g\mu_B$) of linear fit to the plot (Figure 3(d)) is negative (-0.95 ± 0.11), which is consistent with earlier study of ion-irradiation induced sulfur vacancy complexes in ML $MoS_2$[37]. Hence, we attribute the observed e-beam induced sharp peaks to originate from sulfur vacancy complexes. Magneto-optic measurements of $X^{d2}$ and $X^{d3}$ were not performed, and we believe they will be consistent with the literature[37]. A table summarizing assignment of $X^{d1}$, $X^{d2}$, $X^{d3}$ to various defect complexes from previous theory and experiments is presented in SI-XII. Remarkably, the DCP reaches nearly 50% at magnetic field of 9 T. Nonzero DCP at finite magnetic field suggests that the e-beam induced defects peaks involve transition from defect states to pristine valence band, and preserve the spin nature[37]. These spin defects can be explored for quantum sensing applications.

Up to this point, we explored linewidth, localized nature, spatial distribution, spectral spread, charge nature, time stability, and valley Zeeman splitting of e-beam induced sharp defect peaks. The above mentioned characteristics of sharp peaks matches well with previously observed SPEs in ML $MoS_2$[17,18,33,36–38]. Hence, the sharp defect peaks observed in this work are attributed to SPEs.

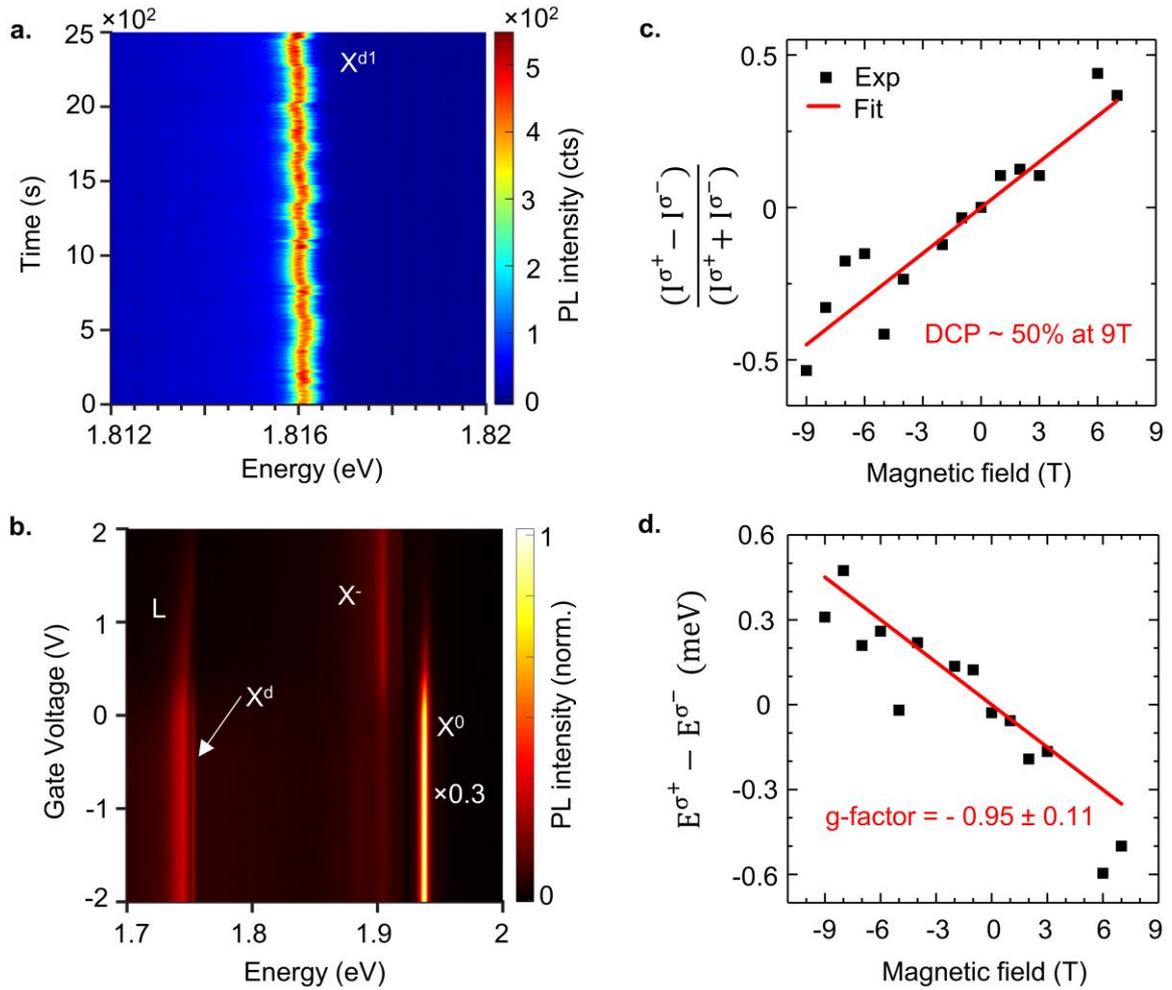

**Figure 3: Time stability, and magneto-optics of sharp defect peak ($X^{d1}$):** **(a)** The spectral jittering (average spectral shift of defect peaks over time) measured for 2500 s (500 spectra, with each spectrum acquired for 5s). **(b)** Gate voltage dependence photoluminescence spectra. L stands for broad defect peak. The $X^0$ peak intensity is divided by factor of 3 for better visibility of other peaks. The degree of circular polarization (DCP = $(I^{\sigma^+} - I^{\sigma^-})/(I^{\sigma^+} + I^{\sigma^-})$), and energy splitting ($E^{\sigma^+} - E^{\sigma^-}$) as a function of magnetic field are plotted in **(c)** and **(d)**, respectively.

**Effect of electron dose and acceleration voltage:**

The density of defects can be tuned by varying electron dose. See SI-I for dose calculation details. Evolution of cryogenic PL spectrum of hBN encapsulated irradiated-ML MoS$_2$ with electron dose are displayed in Figure 4(a). Notably, various hBN thicknesses were used in this dataset, and the thickness of top and bottom hBN can greatly influence the out-coupled PL intensity of ML MoS$_2$[39]. Thus, we only compare normalized PL intensity. In the pristine sample (bottom most graph), a broad low energy peak

(usually referred to as L peak) is observed alongside $X^0$ and $X^-$ peaks. In contrast, $X^d$ SPEs were observed for irradiated samples exposed at low electron doses ($1.57\times10^{13}$ e$^-$/cm$^2$). For higher doses, the $X^d$ peaks broaden due to increase in defect density, and finally become a single broad peak for the highest electron dose ($1.57\times10^{15}$ e$^-$/cm$^2$). Hence, careful optimisation of electron dose can create desired number of defects.

Apart from electron dose, accelerating voltage is another crucial parameter that can affect defect formation. We performed e-beam irradiation at three accelerating voltages (5 kV, 20 kV, 30 kV), keeping the electron dose constant (~$1.57\times10^{13}$ e$^-$/cm$^2$). The corresponding cryogenic PL spectrum of hBN encapsulated irradiated-ML MoS$_2$ is shown in Figure 4(b). We observed multiple sharp defect peaks for 5 kV electron irradiation, whereas higher accelerating voltage (20 kV, 30 kV) irradiations resulted in fewer sharp peaks in a narrow spectral range. In order to understand electron-matter interaction for various acceleration voltages, Monte Carlo simulation of electron trajectories of ML MoS$_2$ on SiO$_2$/Si substrate were performed (SI-XIII) using CASINO software[40]. The observation of multiple sharp peaks for 5 kV irradiation could be due to localized interaction volume near ML and higher probability of backscattering. The backscattered electrons can form additional defects apart from incident electrons[41]. On the other hand, the 30 kV e-beam can penetrate deeper into the substrate with lower backscattering probability, which could result in formation of fewer and highly local defects, as well as defects with higher formation energies. Thus, e-beam accelerating voltage can be a tuning knob for modulating spectral position of the SPEs.

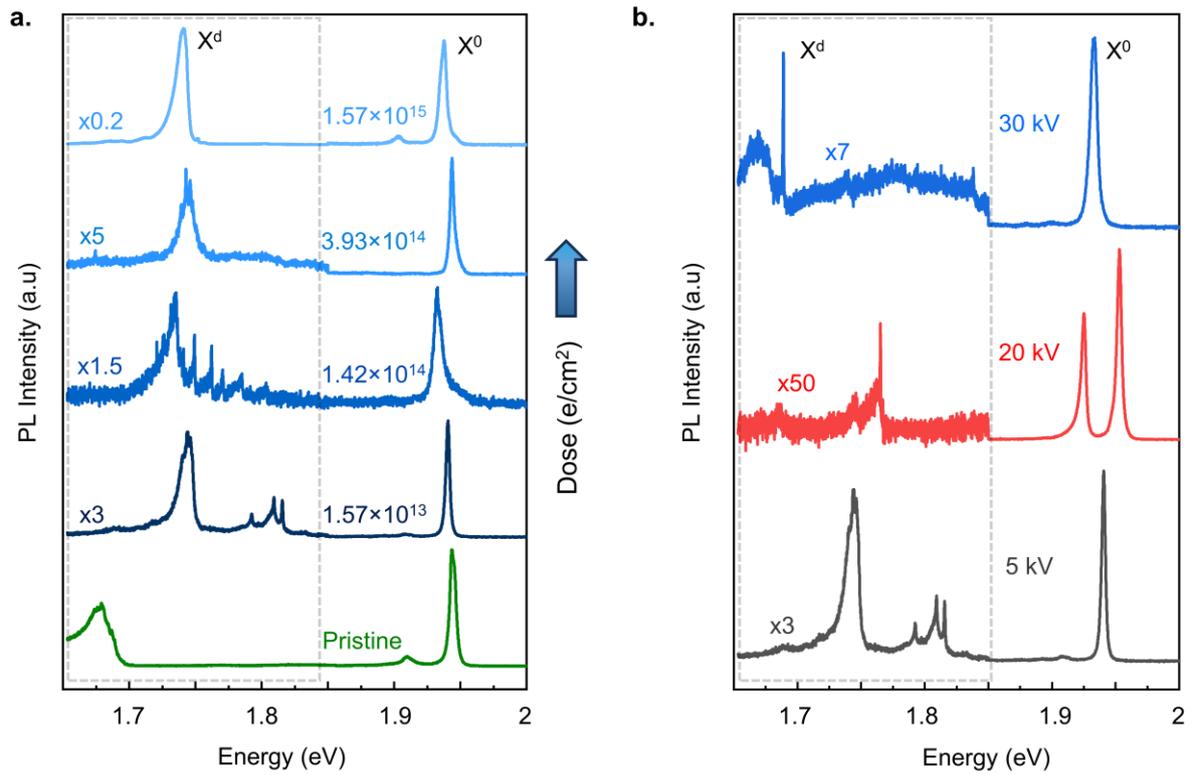

**Figure 4. Evolution of sharp peaks ($X^d$) with electron dose and effect of e-beam accelerating voltage: (a)** Evolution of sharp defect peak PL with electron dose. **(b)** Cryogenic PL of hBN encapsulated ML MoS$_2$ irradiated at 5 kV, 20 kV, and 30 kV e-beam acceleration voltages. All PL spectra are normalized to their maximum value, and defect PL (< 1.85 eV, gray dashed rectangular box) is multiplied by the indicated factors for better visibility.

**Deterministic writing of SPEs:**

Deterministic placement of defects across the ML is crucial for integrating SPEs into cavity structures for practical devices and scaling to large-scale devices. On ML MoS$_2$, deterministic irradiation of varying spot sizes (100 nm, 50 nm, 10 nm) was performed using electron beam lithography (EBL) system (Figure 5(a)). See methods for details of deterministic irradiation. Figure 5(b) shows the optical image of ML MoS$_2$ after irradiation and hBN encapsulation process. Sharp defect peaks were observed in the cryogenic PL spectra from 50 nm and 100 nm irradiated spots (Figure 5(c)). We note that the 10 nm spot region was damaged during the transfer process, and data was not taken on this region. The spatially deterministic placement can be enhanced by using even smaller irradiated spots. Thus, the e-beam irradiation method can be used for the deterministic creation of SPEs with ultra-high positional accuracy and nanometer resolution. We note that the background PL in all of these emitters is very low, indicating high purity for SPEs.

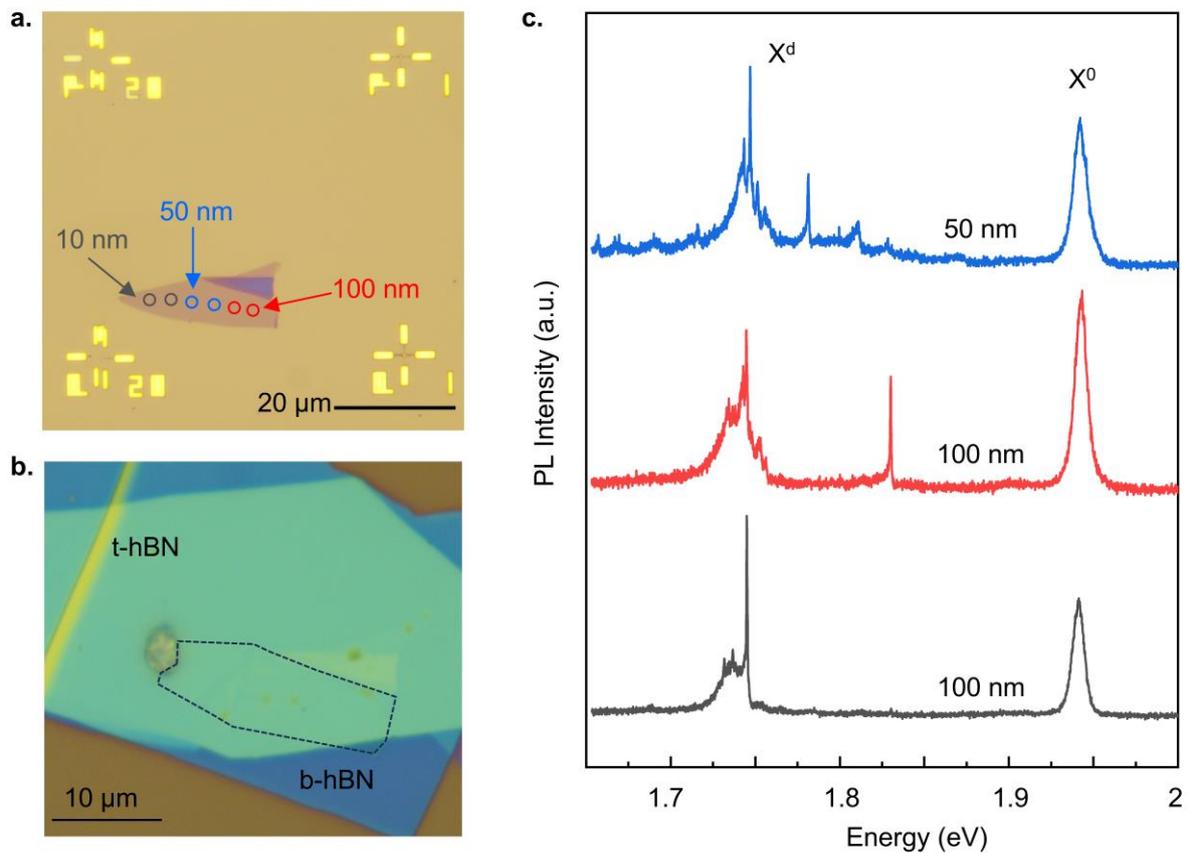

**Figure 5. Deterministic writing of sharp defect peaks with high positioning accuracy via e-beam lithography (EBL) : (a)** Optical image of bare ML $MoS_2$ on $SiO_2$/Si substrates with gold markers for overlay writing in EBL. The overlay irradiation pattern is shown in blue. Hollow circles on ML $MoS_2$ indicate irradiation spots. (Note that the actual irradiation spot is much smaller than the circle) **(b)** Optical image of ML $MoS_2$ (shown in (a)) after e-beam irradiation and hBN encapsulation. **(c)** Cryogenic PL from 50 nm and 100 nm irradiation points.

**Discussion:**

We demonstrated the utilization of ultra-low accelerating voltage e-beam to create SPEs in ML $MoS_2$, with high spatial resolution (< 50 nm, extendable down to 10 nm). We confirmed that the sharp peaks originate from single defects, and have high spectral stability measured by spectral jittering measurements. The sublinear laser power dependence confirmed the defect-bound excitonic nature. Using magneto-optics and PL measurements, we find that SPEs created via e-beam irradiation have similar physical origin to defects created by He ion irradiation[37], attributed to sulfur vacancy complexes[18,42]. The spin nature of SPEs was also confirmed, paving the way towards the development of on-chip quantum communication and sensing devices. The sharp peaks dominate at cryogenic temperature and broaden at higher temperatures due to increasing PSB contributions (decreasing ZPL intensity). Further, we showed that varying electron dose can modify defect density over a large range. On the other hand, e-beam accelerating voltage can tune the spectral position of defect peaks. Lastly, we demonstrated deterministic writing of SPEs using EBL, and observed sharp defect peak emission from irradiated spot sizes down to 50 nm. The SPEs can potentially be positioned with an accuracy of 10 nm, and even better with state-of-the-art EBL systems.

In this work, we have used ML $MoS_2$ as a materials platform for creating SPEs. The e-beam irradiation technique can be generalized to other 2D semiconductors by optimization of electron dose and accelerating voltage. Such ultrahigh spatially deterministic creation of SPEs will enable integration with complex cavity structures including plasmonic structures and photonic cavities. This method will enable next-generation quantum technologies, and can potentially enable fundamental studies of defect-defect coupling and electron-matter interactions.

**Methods:**

**Sample preparation:** $MoS_2$ and hBN flakes are transferred onto $SiO_2$/Si substrates by PDMS-assisted mechanical exfoliation of bulk crystals ($MoS_2$ – 2D semiconductors, hBN – NIMS, Japan). Monolayers (MLs) of $MoS_2$ were identified and confirmed using optical microscopy (RAW imaging[43]). hBN encapsulation of e-beam irradiated samples was performed using the PDMS-PC pick-up technique. For gated device fabrication, graphite/hBN/$MoS_2$/graphite/hBN stack was prepared by PDMS-PC pick-up technique and dropped onto $SiO_2$/Si substrate with alignment markers. It was followed by deposition of gates and contacts of Ti(10 nm)/Au(100 nm) on graphite layers.

**Electron irradiation:** $MoS_2$ MLs were irradiated in a scanning electron microscope (Ultra55 FE-SEM Karl Zeiss) at 5 kV, 20 kV, and 30 kV accelerating voltage. The electron dose was varied by changing magnification and irradiation time with a fixed beam current of 140 pA[28]. For deterministic writing of sharp peaks, ML MoS$_2$ was exfoliated onto SiO$_2$/Si substrate with alignment markers. Point irradiation was performed in an e-beam lithography setup (Raith Pioneer) with a fixed beam current of 180 pA at 5 kV electron accelerating voltage. 100 nm × 100 nm (or 50 nm × 50 nm) boxes were irradiated with separation of 3 μm.

**Cryogenic PL measurements:**

Cryogenic PL was performed in a customized optical set up consisting of Montana cryostation, Andor CCD and spectrometer. 478 nm CW laser was used for excitation. A 50x long-working distance objective with 0.42 numerical aperture was used for focusing laser onto the sample and collection of reflected signal. Appropriate long pass filters were used before spectrometer to block the laser. Cryogenic PL mapping was performed in a custom built optical setup, with piezo scanners inside the cryostat for moving the sample. A 633 nm continuous wave HeNe laser was used for excitation in confocal geometry with microscope objective having numerical aperture 0.82. The laser spot size on the sample is diffraction-limited. For obtaining PL map, a high-resolution grating (1200 lines/mm) was used. A spectral resolution of 40 μeV is measured using the HeNe line. The ML region was scanned with step size of 205 nm and 285 nm in x and y direction, respectively.

**Magneto-optics:**

Magneto optics was performed in Faraday configuration i.e., magnetic field perpendicular to hBN encapsulated irradiated-ML MoS$_2$ atomic layers. A linearly polarized 633 nm laser was used for excitation, and circularly polarized emitted light ($\sigma^+$ and $\sigma^-$) was detected as a function of magnetic field (-9 T to 7 T).

**Author information:**

Corresponding Author: *Akshay Singh, aksy@iisc.ac.in

**Author Contributions:**

AKD and AS developed the experimental framework. AKD performed electron irradiation, sample fabrication, and optical characterisation with assistance from MPS and SKY. MPS performed e-beam lithography, with guidance from AN. SR and CR performed cryogenic PL mapping and magneto optics experiments, with guidance from XM. AKD performed the data analysis, with assistance in PL data analysis by SR and SKY. KW and TT provided the hBN bulk crystals. AKD, SKY and AS discussed and prepared the manuscript, with contributions from all authors.

**Data Availability**

All data is available upon reasonable request.

**Acknowledgments**

AS would like to acknowledge funding from Indian Institute of Science start-up grant, SERB grant (SRG-2020-000133), and DST Nanomission CONCEPT (Consortium for Collective and Engineered Phenomena in Topology) grant. We acknowledge financial support from the Quantum Research Park, which is funded by the Government of Karnataka, India. AKD would like to acknowledge Prime Minister's Research Fellowship (PMRF). The authors also acknowledge Micro Nano Characterization Facility (MNCF), Centre for Nano Science and Engineering (CeNSE) for use of characterization facilities. SKY would like to acknowledge the assistance received from SERB (PDF/2023/000658). Part of this work was supported by the Agence Nationale de la Recherche under the program ESR/EquipEx+ (Grant No. ANR-21-ESRE- 0025) and ANR projects ATOEMS and IXTASE.

# Supplementary Information

**Quantum light generation with ultra-high spatial resolution in 2D semiconductors via ultra-low energy electron irradiation**

Ajit Kumar Dash[1], Sharad Kumar Yadav[1], Sebastien Roux[2], Manavendra Pratap Singh[3], Kenji Watanabe[4], Takashi Taniguchi[5], Akshay Naik[3], Cedric Robert[2], Xavier Marie[2], Akshay Singh[1*]

[1]Department of Physics, Indian Institute of Science, Bengaluru, Karnataka 560012, India

[2]Université de Toulouse, INSA-CNRS-UPS, LPCNO, 135 Avenue de Rangueil, 31077 Toulouse, France

[3]Centre for Nanoscience and Engineering, Indian Institute of Science, Bengaluru, Karnataka 560012, India

[4]Research Center for Functional Materials, National Institute for Materials Science, Japan

[5]International Center for Materials Nanoarchitectonics, National Institute for Materials Science, Japan

*Corresponding author: aksy@iisc.ac.in

**SI-I Methods of varying electron dose in SEM**

**SI-II Voigt curve fitting and observed splitting of neutral exciton peak ($X^0$).**

**SI-III: Environmental passivation of electron beam induced defects.**

**SI-IV: Long-term stability of sharp $X^d$ peaks.**

**SI-V: Effect of annealing in $N_2$ environment.**

**SI-VI: Laser power dependent PL.**

**SI-VII: Procedure for obtaining filtered cryogenic PL map.**

**SI-VIII: Cryogenic PL spectrum of additional irradiated samples.**

**SI-IX: Asymmetric nature of sharp defect peaks attributed to phonon side bands.**

**SI-X: Gate voltage dependent PL.**

**SI-XI: Raw PL spectra of $X^{d1}$ peak at -6 T, 0 T, and 6 T magnetic field.**

**SI-XII: Assignment of defect peaks from literature.**

**SI-XIII: Monte-Carlo simulation of electron trajectories.**

**SI-I Methods of varying electron dose in Scanning Electron Microscope (SEM)**

The density of defects created solely depends on the electron dose received by the sample. In a SEM, the dose can be varied by changing beam current, exposure time, and irradiation area[1]. We have found that changing beam current can modify the beam size, that in turn can modify the defect creation in a complex manner. Thus, in this work, the dose is varied by adjusting exposure time and irradiation area (magnification), at a fixed beam current of 140 pA. The electron dose received by the sample can be calculated by the formula, $e^- \, dose = \frac{I \times t}{e \times A}$, where $I, A$ stand for beam current and area scanned on the sample, respectively, $t$ stands for total irradiation time, and $e$ is the electronic charge. The minimum electron dose used in this work (~$1.57 \times 10^{13}$ e⁻/cm²) is comparable to the helium ion dose (~$2 \times 10^{12}$ ions/cm²) used to obtain sharp defect peaks in ML $MoS_2$[2]. However, the momentum carried by e-beam is around 7,400 times less than ion beam. Hence, using e-beam to create sharp defect emission centres is expected to cause less damage to the surrounding lattice than ion beam irradiation, potentially resulting in SPEs with higher brightness and higher purity.

**SI-II: Voigt curve fitting and observed splitting of neutral exciton peak (X⁰)**

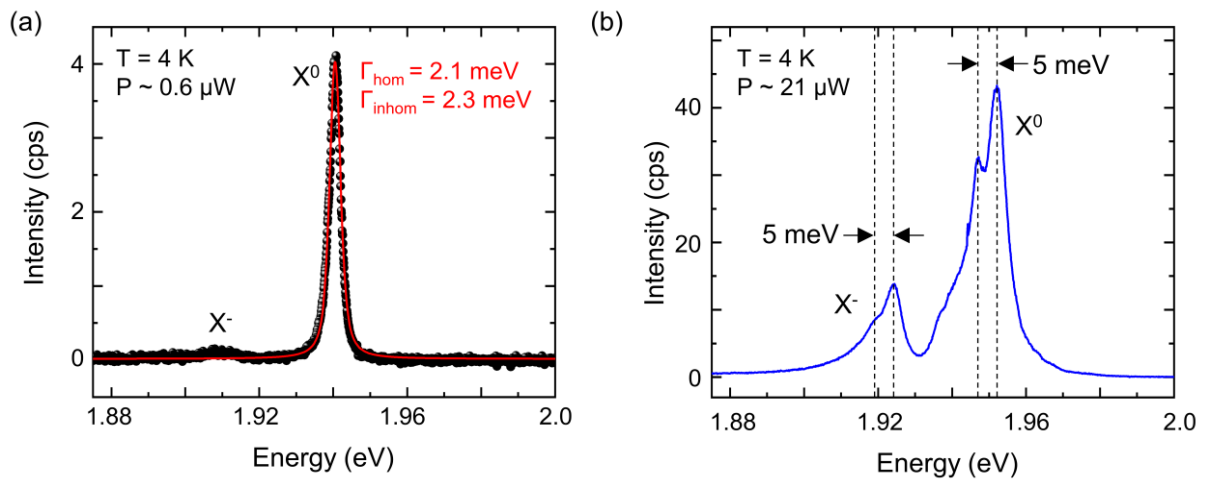

**Figure S1**: **(a)** The Voigt function fitting of neutral exciton (X⁰) peak of hBN encapsulated irradiated-monolayer

(ML) MoS$_2$ sample (S1). The extracted homogeneous ($\Gamma_{hom}$, Lorentzian component) and inhomogeneous ($\Gamma_{inhom}$, Gaussian component) linewidths are 2.1 and 2.3 meV, respectively. **(b)** Cryogenic PL spectrum from another location of S1 sample showing splitting of neutral exciton (X$^0$) and trion (X$^-$) peaks. The observed splitting in the PL spectrum is approximately 5 meV. The splitting ranges from 5 – 9 meV in other irradiated samples.

## SI-III: Environmental passivation of electron beam induced defects

In order to study environmental passivation of e-beam induced defects, two freshly exfoliated ML MoS$_2$ flakes were irradiated at same conditions as S1. One of the irradiated MLs was hBN encapsulated within 30 minutes of electron irradiation, and the other one after 24 hours of electron irradiation. We observed sharp defect peaks for ML MoS$_2$ whose encapsulation was performed quicky after electron irradiation (Figure S2a). However, sharp defect PL was quenched for ML MoS$_2$ whose encapsulation was performed after 24 hours of irradiation (Figure S2b). Hence, the e-beam induced defects get quickly passivated by environmental gases and hydrocarbons, which is also evidenced by previous work on MoS$_2$ defects created by proton beam irradiation[3]. The filling of defects may be completely avoided by attaching a viscoelastic transfer system to the SEM chamber, allowing hBN encapsulation without exposure to the environment.

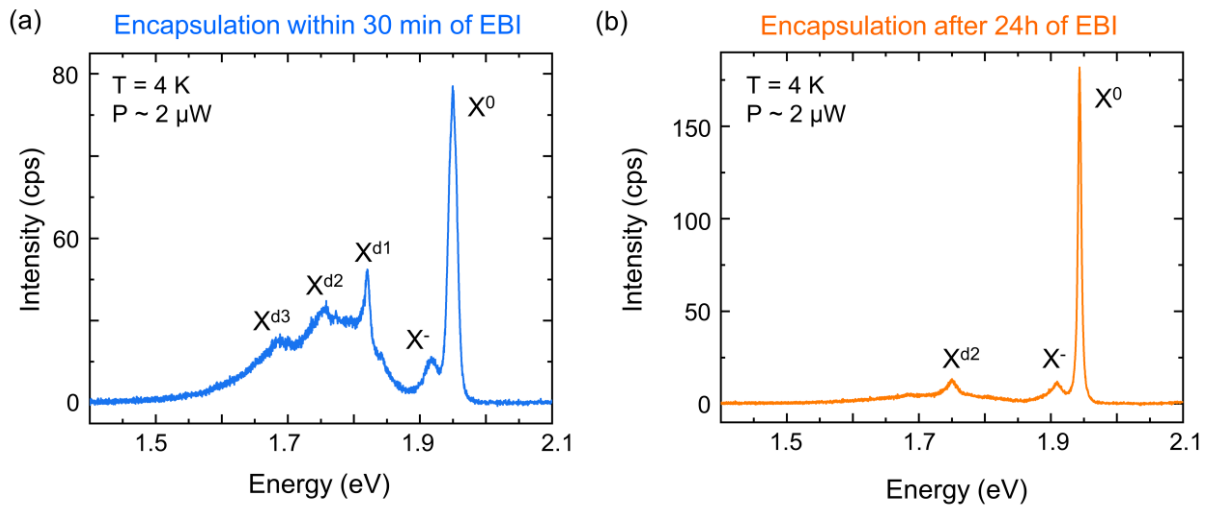

**Figure S2**: Cryogenic PL spectra of ML MoS$_2$ samples that are hBN encapsulated within 30 minutes **(a)** and after 24 hours **(b)** of electron beam irradiation. The passivation of optically active defects by environmental gases and hydrocarbons results in PL quenching of sharp defect peaks (X$^d$).

## SI-IV: Long-term stability of sharp $X^d$ peaks

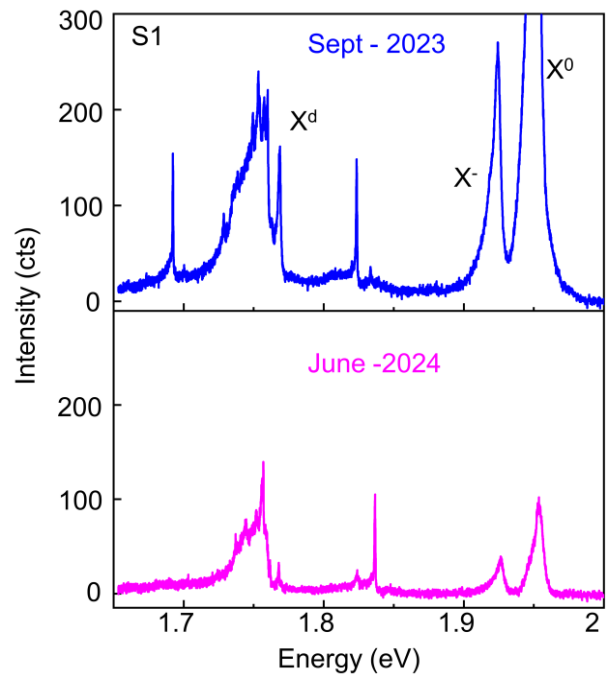

**Figure S3:** Cryogenic PL spectra of fresh S1 sample (top panel) and taken after ten months (bottom panel). Sharp $X^d$ peaks remain after multiple cooling cycles, PL, and magneto-optics experiments.

## SI-V: Effect of annealing in $N_2$ environment

Annealing of 2D material based heterostructure in vacuum/inert atmosphere is widely used for removing the trapped contaminants, which in turn improves the interface quality[4]. However, the annealing process may introduce additional defects if the process is unoptimized[5]. In order to understand effect of annealing on sharp defect peaks, a hBN encapsulated irradiated-ML $MoS_2$ was annealed at 250 $^0$C for 3 hours in $N_2$ filled glovebox (annealing conditions generally used for improving heterostructure interface quality[1]). Irradiation conditions were kept the same as samples S1, S2, S3. We observed broader (FWHM ~ tens of meV) defect peaks instead of sharp defect peaks (FWHM ~ 500 μeV) in annealed samples (Figure S4), which indicates that additional defects are introduced by annealing process. We note that the annealing process could be optimized for improving SPE emission, but we have not attempted this optimization.

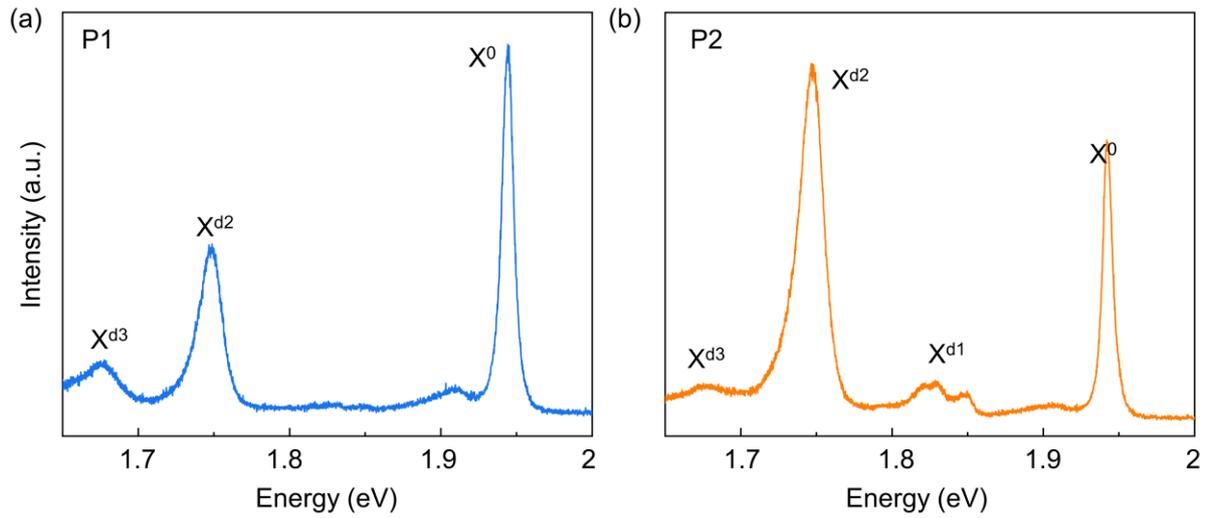

**Figure S4**: **(a, b)** Cryogenic PL spectrum of annealed hBN encapsulated ML MoS$_2$ (irradiated at same conditions as S1). Spectra are normalized to maximum for better visibility. P1 and P2 correspond to different points of the same sample.

## SI-VI: Laser power dependent PL

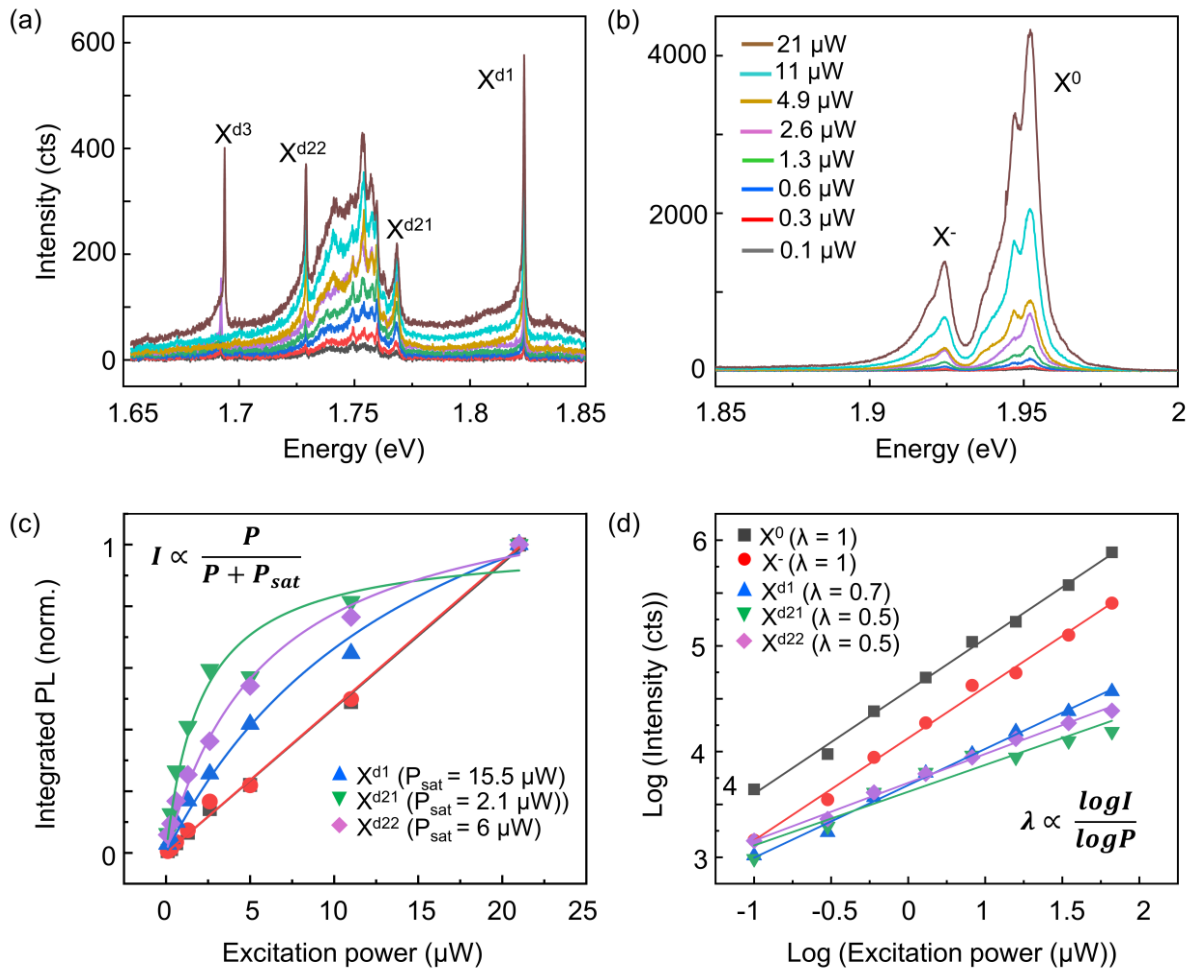

**Figure S5**: **(a, b)** Laser power dependent cryogenic PL spectra of hBN encapsulated irradiated-ML MoS$_2$ sample (S1). X$^0$, X$^-$, X$^d$ stands for neutral, charged, and bound exciton peaks, respectively. **(c)** Saturation powers (P$_{sat}$) obtained by fitting power law (P = P/(P+P$_{sat}$)). **(d)** Laser power coefficients (λ) of X$^0$, X$^-$, and all X$^d$ peaks.

## SI-VII: Procedure for obtaining filtered cryogenic PL map

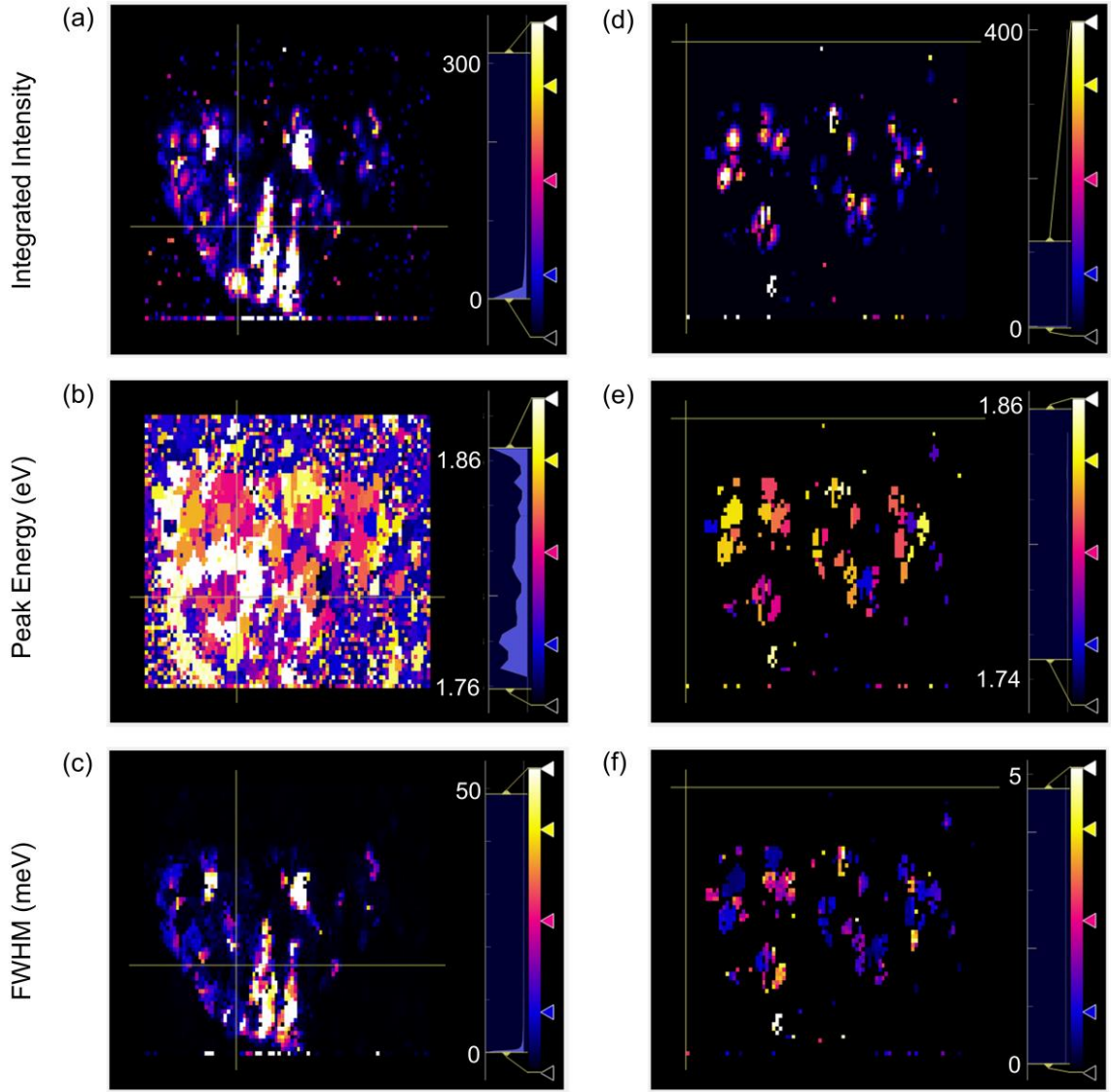

**Figure S6**: Unfiltered **(a-c)** and filtered **(d-f)** maps of integrated intensity, peak energy, and FWHM. The filtering method is described in the text below.

A Lorentzian fit is performed on each spectra, regardless of whether a sharp X$^d$ peak appears or not. The data are filtered by R-squared value [0.6 to 1] (to keep only good quality fits), peak energy [1.77 eV to 1.86 eV] (to remove edge effects) and FWHM [0.3 meV to 5 meV] (to measure the sharp X$^d$ peak, remove random spikes due to cosmic rays and remove broad emission from the background). Only the fits that satisfy these three conditions are considered in the corresponding maps. Figure S6 compares unfiltered and filtered maps integrated intensity, peak energy, and FWHM.

**SI-VIII: Cryogenic PL spectrum of additional irradiated samples**

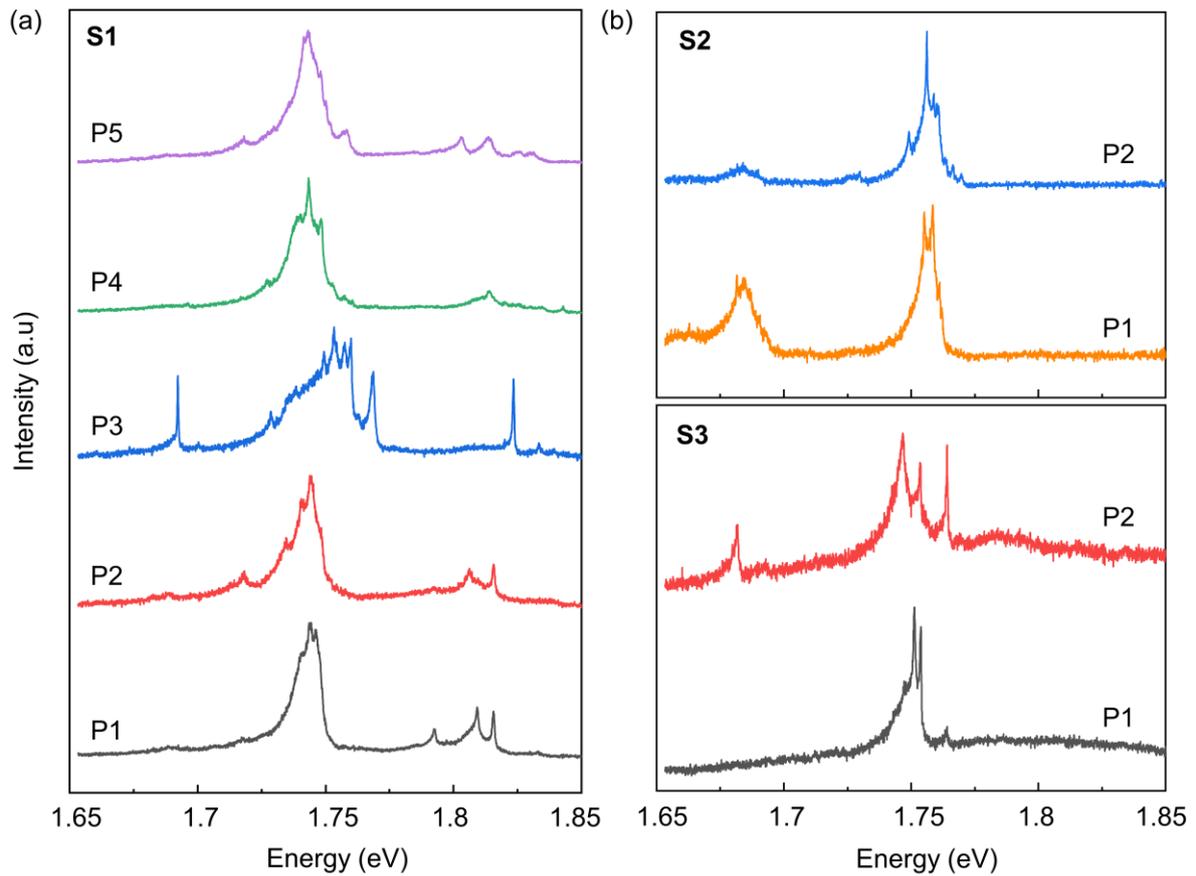

**Figure S7**: **(a)** Defect PL in extended spectral range across different points of S1 sample. **(b)** Cryogenic PL spectrum of additional hBN encapsulated ML $MoS_2$ samples (S2, S3) irradiated at same conditions as S1. All spectra were acquired using 478 nm laser excitation and at average power of ~ 2.5 µW. Spectra are normalized to maximum (in each spectra) for better visibility.

## SI-IX: Asymmetric nature of sharp defect peaks attributed to phonon side bands

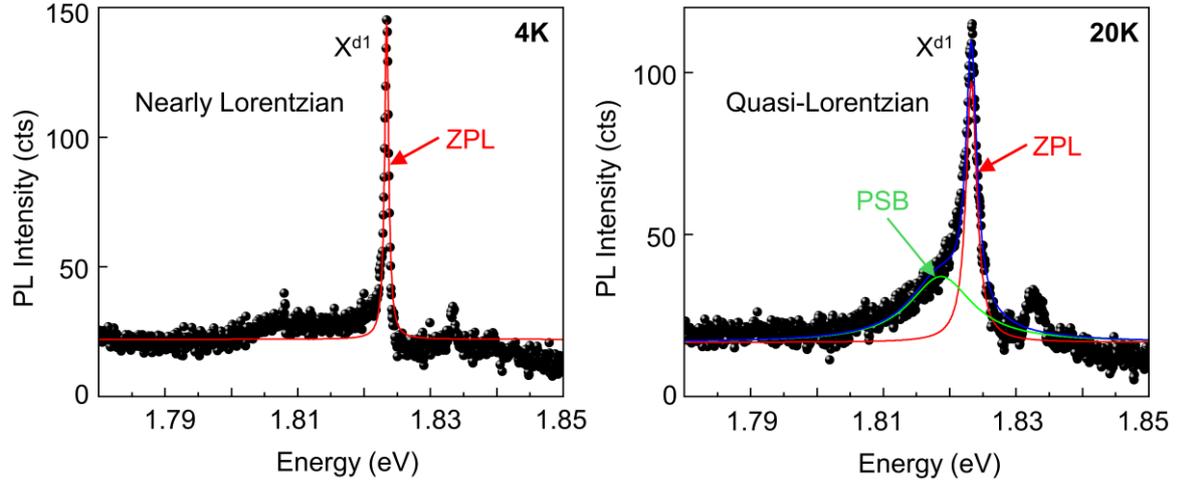

**Figure S8**: PL spectrum of sharp defect peak ($X^{d1}$) at 4 K and 20 K temperatures. Lorentzian fittings for both spectra are shown as solid lines. Here, ZPL and PSB stands for zero phonon line and phonon side band, respectively.

## SI-X: Gate voltage dependent PL

The gated device can be considered as a parallel plate capacitor with bottom graphite and 1L MoS$_2$ as plates and bottom hBN as dielectric between the plates. The dielectric constant of hBN $\varepsilon_{hBN} = 2.5$. The thickness of the hBN ($d$) is 20 nm as confirmed by AFM. The charge carrier density can be obtained from the relation $n = \varepsilon_0 \varepsilon_{hBN} V/ed$, where $\varepsilon_0$, $V$, and $e$ are absolute perimittivity of free space, applied gate voltage, and electronic charge, respectively. The calculated charge carrier density for 20 nm thick hBN and 2 Volts applied gate voltage would be $\sim 10^{12}$ per cm$^2$.

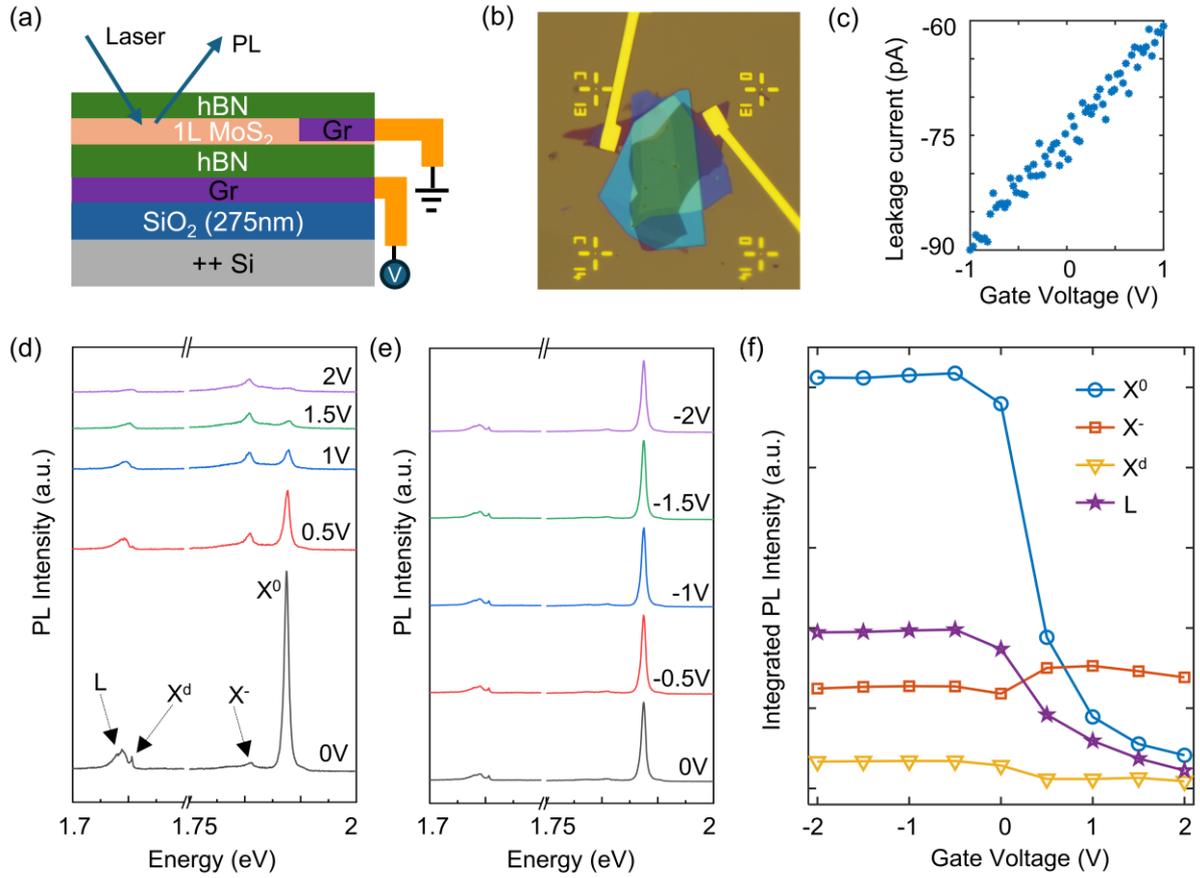

**Figure S9:** Gate voltage dependent PL of hBN encapsulated irradiated-ML MoS$_2$. The schematics and optical image of the gated device are shown in **(a)** and **(b)**, respectively. **(c)** The measured leakage current of the device at room temperature. The evolution of cryogenic PL of the sample with applied positive (electron doping) and negative (hole doping) gate voltage are plotted in **(d)** and **(e)**, respectively. **(f)** The variation of integrated PL intensity of neutral exciton ($X^0$), trion ($X^-$), sharp defect peak ($X^d$), and L-peak with gate voltage.

## SI-XI: Raw PL spectra of $X^{d1}$ peak at -6 T, 0 T, and 6 T magnetic field

We performed magneto-PL with magnetic field oriented perpendicular to atomic layers (Faraday geometry). Circularly polarized PL ($\sigma^+$ and $\sigma^-$) was detected after excitation with linearly polarized laser. Raw spectra of $\sigma^+$ and $\sigma^-$ emitted intensity of $X^{d1}$ peak at 0 T, 6 T, and -6 T magnetic field are shown in Figure S10.

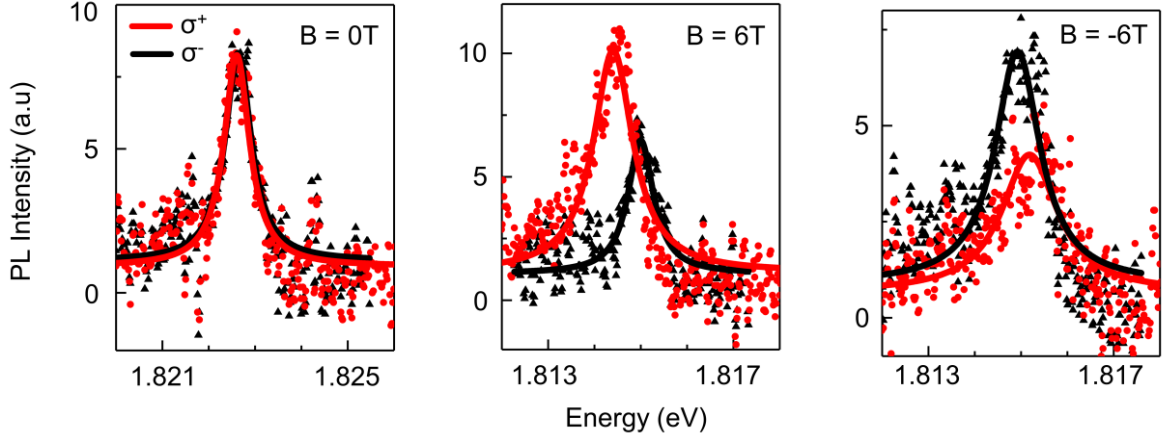

**Figure S10**: Raw spectra of σ⁺ and σ⁻ emitted intensity of $X^{d1}$ peak at 0 T, 6 T, and -6 T magnetic fields.

## SI-XII: Assignment of defect peaks from literature

As mentioned earlier, there are three spectral regimes where sharp peaks appear, i.e., $X^{d1}$, $X^{d2}$, $X^{d3}$. The PL spectra across different points of the S1 sample in the extended spectral range, and of two additional samples irradiated at the same condition as S1, are presented in SI-V. The three spectral regimes correspond to various types of defect complexes[6]. For example, in a recent study[7], sulfur vacancies ($V_s$) and charged oxygen ad atoms ($O_{ad}$) were shown to be the most prevalent defects in exfoliated ML $MoS_2$ due to their lower formation energies compared to molybdenum vacancies ($V_{Mo}$). Table 1 summarizes assignment of defect peaks in ML $MoS_2$ created by various methods by both theory and experiment.

|  | **Method of defect Creation** | **Position below $X^0$ peak (meV)** | **Defect peak assignment** |
| --- | --- | --- | --- |
| **Klein et. al[2]** | Helium ion irradiation | 100 - 220 | $V_{Mo}$ (DFT) |
| **Wang et. al[6]** | UV irradiation | 50 - 300 | $V_s$, $V_{2s}$, $S_{ad}$ (DFT) |
| **Kumar et. al[7]** | Chemical Vapour Deposition (CVD) | 100 - 300 | $V_s$, $S_{ad}$, $O_{ad}$ (DFT) |

**Table 1**: Assignment of defect PL peaks in ML $MoS_2$ from literature. $V_{Mo}$, $V_s$, $V_{2s}$, $S_{ad}$, $O_{ad}$ stands for molybdenum vacancy, sulfur vacancy, disulfur vacancy, sulfur ad atom, and oxygen ad atom, respectively.

## SI-XIII: Monte-Carlo simulation of electron trajectories

In order to study the interaction of electron beam with ML $MoS_2$ on $SiO_2$/Si substrate, Monte-Carlo simulation of trajectories of primary and back-scattered electrons were performed using CASINO software. An electron spot size of 1 nm was used in the microscope set up. The trajectories of primary and back-scattered electron beam for 5 kV, and 30 kV acceleration voltage are plotted in Fig. S11 (a), and (b), respectively. At 5 kV, the electron interaction volume is localized near ML $MoS_2$, and the probability of backscattering of electrons is high. The backscattered electrons can form additional defects apart from incident electrons, which could explain the observation of multiple sharp peaks at various sample locations. Contrastingly, the electron can penetrate deeper into the substrate at higher accelerating voltages (30 kV), and the probability of backscattering is less than 5 kV. Hence, higher acceleration voltages can create highly local and fewer defects. We note that more studies need to be performed, keeping hBN thicknesses same, to find out whether emission intensities are also affected by the accelerating voltage. We also note that broad beam irradiation was used for Figures 1-4 of main text, and point irradiation was used for Figure 5 of main text.

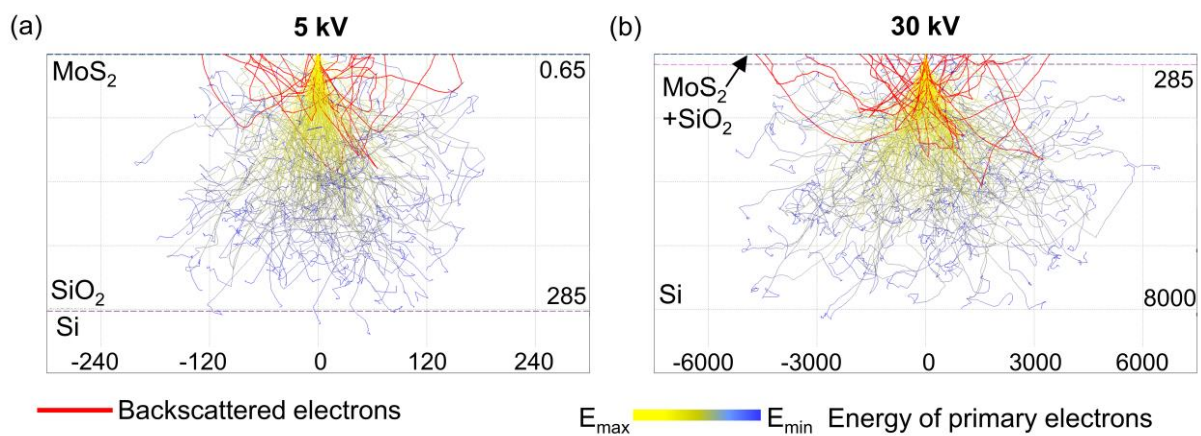

**Figure S11**: Monte-Carlo simulation of electron trajectories of ML $MoS_2$ on $SiO_2$/Si substrate for **(a)** 5 kV, and **(b)** 30 kV electron accelerating voltage. The thickness of ML $MoS_2$, $SiO_2$, and Si used in the simulation are 0.65 nm, 285 nm, and 500 μm, respectively. The unit of x and y axis are in nm.